\documentclass[%
 reprint,
superscriptaddress,
 amsmath,amssymb,
 aps,
 pre,
floatfix,
]{revtex4-2}
\usepackage{cancel}
\usepackage[caption=false]{subfig}
\usepackage{graphicx}
\usepackage{dcolumn}
\usepackage{bm}
\usepackage[hidelinks]{hyperref}
\usepackage{xcolor}
\usepackage[normalem]{ulem}
\definecolor{ra}{rgb}{0.8, 0.0, 0.0}

\begin{document}

\preprint{APS/123-QED}

\title{Lattice $\phi^{4}$ field theory as a multi-agent system of financial markets}

\author{Dimitrios Bachtis}
\email{d.bachtis@qmul.ac.uk}
\affiliation{Centre for Theoretical Physics, Department of Physics and Astronomy,
Queen Mary University of London, London E1 4NS, United Kingdom}

\include{ms.bib}

\date{November 25, 2024}

\begin{abstract}

We introduce a $\phi^{4}$ lattice field theory with frustrated dynamics as a multi-agent system to reproduce stylized facts of financial markets such as fat-tailed distributions of returns and clustered volatility. Each lattice site, represented by a continuous degree of freedom, corresponds to an agent experiencing a set of competing interactions which influence its decision to buy or sell a given stock. These interactions comprise a cooperative term, which signifies that the agent should imitate the behavior of its neighbors, and a fictitious field, which compels the agent instead to conform with the opinion of the majority or the minority. To introduce the competing dynamics we exploit the Markov field structure to pursue a constructive decomposition of the $\phi^{4}$ probability distribution which we recompose with a Ferrenberg-Swendsen acceptance or rejection sampling step. We then verify numerically that the multi-agent $\phi^{4}$ field theory produces behavior observed on empirical data from the FTSE 100 London Stock Exchange index. We conclude by discussing how the presence of continuous degrees of freedom within the $\phi^{4}$ lattice field theory enables a representational capacity beyond that possible with multi-agent systems derived from Ising models. 
\end{abstract}

\maketitle

\section{\label{sec:level1}Introduction} Statistical physics has been extensively utilized as a mathematical framework to contribute towards advancing quantitative finance~\citep{bpbook}. One reason which motivates the implementation of statistical-mechanical models for the study of financial problems is the observation that financial prices are governed by universal behavior~\citep{mand,Fama,Mantegna1995,Ghashghaie1996,GALLUCCIO1997423} which is reminiscent of that encountered as an emergent phenomenon in interacting many-body problems of statistical physics. Various studies have therefore appeared, which aim to formulate a statistical-mechanical framework capable of recovering this universal behavior as a consequence of the interaction between agents which aim to buy or sell a given stock~\citep{Lux1999,bornholdt,Giardina2003,Zhou2007,BAK1997430,bougiamez,levyjpa}. 

Such studies usually consider two distinct types of agents, called fundamentalists or chartists. Fundamentalists form decisions based on the fundamental value of an asset: they are inclined to buy a stock when they consider it undervalued or sell a stock if they believe it resides above the fundamental value. Chartists do not take into consideration the fundamental value of an asset but form decisions using information obtained from trends, or the behavior of others. The coexistence of such agents then gives rise to herding behavior~\citep{Cont_Bouchaud_2000,bouchaud1998,corcosetal}.

Spin models in statistical physics provide a starting point for the simulation of systems comprising interacting agents. For example, one can directly map the decision of an agent to buy or sell a stock on the binary degrees of freedom of the Ising model. The phase transition of the system can then provide insights on the cooperative dynamics which force agents to collectively buy or sell a stock. Agents in the context of financial markets aim to maximize their future returns and this is additionally dependent on whether they share the opinion of the majority or the minority~\citep{challet}. Spin models can then be further enhanced by additional terms, see for instance Refs.~\citep{bornholdt,KAIZOJI2002441,krausebo,taka1,taka2}, to take into consideration this type of interaction. Ising and Potts models, as well as their disordered variants, have been historically successful as multi-agent models of financial markets~\citep{Bouchaud2013,Sornette_2014}. 

In this manuscript, we introduce the $\phi^{4}$ lattice scalar field theory as a multi-agent system to describe buy or sell dynamics of interacting agents in a model of financial markets where expectation bubbles and crashes emerge. Each degree of freedom, represented by a continuous value, corresponds to the decision of an agent to buy or sell a given stock. The agent is then experiencing a set of competing interactions.

The first interaction is a cooperative term, already present in the lattice action of the $\phi^{4}$ theory, which signifies that each agent should imitate the decision of its neighbors. The second interaction, which must be introduced explicitly in the lattice action by being coupled to a fictitious field, is a compulsion term which compels an agent to conform with the opinion of the majority or minority. We then discuss how the $\phi^{4}$ theory has sufficient representational capacity to study the presence of fundamentalists and chartists in a model of financial markets where expectation bubbles or crashes emerge.

The dynamics discussed herein can only be introduced within a Gibbs sampling approach~\citep{GemanGeman}. We discuss how the Markov property of the $\phi^{4}$ theory~\citep{PhysRevD.103.074510,Bachtis_2022,Bachtis:202274} enables the incorporation of Gibbs sampling for the system. After pursuing a constructive decomposition of the probability distribution in order to introduce the competing interactions, we investigate if we can recompose it with a Ferrenberg-Swendsen acceptance or rejection sampling step~\citep{PhysRevLett.63.1195}.

\begin{figure}[t]
\includegraphics[width=6.5cm]{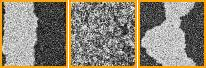}
\caption{\label{fig:fig1} Configurations of the multi-agent $\phi^{4}$ theory for $20500$, $28311$, and $30000$ Monte Carlo sweeps from left to right, respectively. The configurations are obtained for lattice size $L=64$ and for couplings $m^{2}=-3.0$, $\lambda=0.7$, and $a=5.0$.  White color corresponds to the largest positive value of the fields, black color to the largest negative value, and shades of grey represent the intermediate values. We observe transitions from a metastable state of $20500$ sweeps, to a phase of chaotic dynamics in $28311$ sweeps, and a subsequent metastable state of $30000$ sweeps. Chartists dominate in the state of chaotic dynamics, and metastable states indicate the emergence of an expectation bubble.}
\end{figure}

We then utilize the multi-agent $\phi^{4}$ theory to explore if we can observe the emergence of stylized facts of financial markets such as fat-tailed distributions of returns and clustered volatility. We compare the results against empirical data obtained from the FTSE 100 London Stock Exchange index to discuss if the multi-agent $\phi^{4}$ theory could be extended on quantitative problems of modeling empirical financial data, for example via the solution of the inverse problem~\citep{PhysRevD.103.074510}. We additionally investigate if the $\phi^{4}$ theory can reproduce critical exponents consistent with empirical studies. We conclude by discussing how the presence of continuous degrees of freedom within the $\phi^{4}$ lattice field theory enables a representational capacity beyond that possible with multi-agent systems derived from Ising models.

\section{\label{sec:level2} Constructive decomposition of the  $\phi^{4}$ theory} The $\phi^{4}$ scalar field theory is described by the lattice action:
\begin{equation}\label{eq:oldaction}
S=-J\sum_{\langle ij \rangle}  \phi_{i} \phi_{j} + \bigg(d+\frac{m^{2}}{2}\bigg) \sum_{i} \phi_{i}^{2}+\frac{\lambda}{4} \sum_{i} \phi_{i}^{4},
\end{equation}
where $J=1$ is a ferromagnetic coupling, $d=2$ is the dimension of the system, $m^{2}$ and $\lambda$ are the squared mass and lambda couplings, and $\langle ij \rangle$ defines a nearest-neighbor interaction on a square lattice.  For a given $\lambda>0$ the system undergoes a second-order phase transition for a unique value $\mu^{2}<0$ due to a spontaneous breaking of the $Z_{2}$ symmetry. In the limit $\lambda \rightarrow \infty$ and $m \rightarrow -\infty$ the system reduces to the Ising model.

 We consider that the system is simulated with helical boundary conditions on a lattice volume $V=L\times L$, where $L$ is the lattice size, and that it is described by the Boltzmann probability distribution:
\begin{equation}\label{eq:origprob}
p_{\Phi_{i}}= \frac{\exp[-S_{\Phi_{i}}]}{Z},
\end{equation}
where $\Phi_{i}$ denotes a given configuration and $Z=\int_{-\infty}^{\infty} \exp[-S_{\Phi}] d\Phi$ is the partition function of the system.

In order to study numerically a multi-agent version of the $\phi^{4}$ quantum field theory it is necessary to pursue a constructive decomposition of its probability distribution. The first decomposition is to enable Gibbs sampling~\citep{GemanGeman} via conditional probability distributions since the compulsion term can be incorporated only under the constraint of a heat-bath approach. The second decomposition emerges from the necessity to sample the quartic polynomial lattice action $\phi^{4}$ term in order to recompose the probability distribution of the $\phi^{4}$ theory,  a problem that we solve via the incorporation of a Ferrenberg-Swendsen acceptance or rejection sampling step~\citep{PhysRevLett.63.1195}. To simplify the derivations as much as possible we first discuss the case of the conventional $\phi^{4}$ theory of Eq.~\ref{eq:oldaction} and gradually  introduce the concepts. 

The two-dimensional $\phi^{4}$ theory on a square lattice is a Markov random field~\citep{PhysRevD.103.074510}. By considering that the lattice sites form a set $\Lambda$, then the $\phi^{4}$ theory satisfies the local Markov property which enables Gibbs sampling of a conditional probability distribution as:
\begin{align*}
& p(\phi_{i}|\phi_{j,j\in \Lambda-i})= p(\phi_{i}|\phi_{j,j\in n_{i}})= \\ & \frac{\exp\Big[\phi_{i}\sum_{j,j\in n_{i}}\phi_{j}-(2+m^{2}/2)\phi_{i}^{2}-(\lambda/4) \phi_{i}^{4}  \Big]}{\int_{-\infty}^{\infty}\exp\Big[\phi_{i}\sum_{j,j\in n_{i}}\phi_{j}-(2+m^{2}/2)\phi_{i}^{2}-(\lambda/4) \phi_{i}^{4}  \Big] d\phi_{i}},
\end{align*}
where $n_{i}$ denotes the set of neighbors of lattice site $i$. 

We define $c_{1}=\sum_{j,j\in n_{i}}\phi_{j}$, $c_{2}=(2+m^{2}/2)$, $c_{3}=\lambda/4$, and express the conditional probability distribution as:
\begin{equation}
p(\phi_{i}|\phi_{j,j\in n_{i}})= \frac{f_{1}(\phi_{i})f_{2}(\phi_{i})}{\int_{-\infty}^{\infty}f_{1}(\phi_{i})f_{2}(\phi_{i})d\phi_{i}},
\end{equation}
where:
\begin{equation}
\frac{f_{1}(\phi_{i})}{\int_{-\infty}^{\infty}f_{1}(\phi_{i})d\phi_{i}}=\frac{\exp[-c_{2}(\phi_{i}-c_{1}/2c_{2})^{2}]}{\sqrt{\pi/c_{2}}},
\end{equation}
\begin{equation}
f_{2}(\phi_{i})=\exp[-c_{3}\phi_{i}^{4}].
\end{equation}

The function $f_{1}(\phi_{i})$ corresponds to a Gaussian distribution $\mathcal{N}(\mu,\sigma^{2})$ with mean $\mu$ and standard deviation $\sigma$:
\begin{equation}
\label{eq:mu}
\mu=\frac{c_{1}}{2c_{2}}, \ \sigma=\frac{1}{\sqrt{2c_{2}}}.
\end{equation}

The conditional probability distribution $p(\phi_{i}|\phi_{j,j\in n_{i}})$ of the $\phi^{4}$ theory can then be recomposed with a Ferrenberg-Swendsen acceptance or rejection sampling step. Specifically, one first samples a proposed degree of freedom $\phi_{i}'$ from the Gaussian probability distribution $\mathcal{N}(\mu,\sigma^{2})$ which corresponds to the $f_{1}(\phi_{i})$ function. One then draws a random number $u$ from the uniform probability distribution $U[0,1)$ and calculates the function $f_{2}(\phi'_{i})$ with the already sampled degree of freedom $\phi'_{i}$. If $f_{2}(\phi'_{i})>u$ we then accept the proposed degree of freedom $\phi'_{i}$ as valid in the Monte Carlo approach. This recomposition of the probability distribution is possible only for $c_{2}>0$ and for sufficiently small values of $c_{3}$. Using the aforementioned framework one is capable of sampling the probability distribution of the conventional $\phi^{4}$ theory. We are now interested in extending these concepts to the sampling of a multi-agent $\phi^{4}$ theory which acts as a model for interacting agents in the context of financial markets.

\section{\label{sec:level3} Multi-agent $\phi^{4}$ theory as a model of financial markets} A simplified framework to describe dynamics of interacting agents in the context of financial markets is expected to incorporate at least two interactions between agents. 

The first interaction should take into consideration that agents tend to imitate their neighbors when deciding to sell or buy a certain stock, therefore giving rise to herding behavior. This type of interaction is already present in the lattice action of Eq.~\ref{eq:oldaction}, due to the $Z_{2}$ symmetry-breaking phase transition. Via the $Z_{2}$ symmetry-breaking degrees of freedom will be collectively aligned to a positive or a negative value which encodes a collective decision to buy or sell a stock, respectively.

The second type of interaction should incorporate the preference of an agent to be part of the opinion of the majority or minority. For instance, consider the possibility that an agent, such as a fundamentalist, who is part of the majority might desire to align with the opinion of the minority in order to evade a potential crash of a given overvalued stock. Alternatively, a fundamentalist agent might desire to share the opinion of the minority if they believe that a certain stock is undervalued so that the agent could benefit from future returns. In the case of chartists, an agent in the minority might decide to spontaneously transition to the opinion of the majority for immediate gains due to an emerging trend. This type of interaction is not present in the lattice action of Eq.~\ref{eq:oldaction} and must be explicitly introduced. 

We remark that the magnetization is a sum over the degrees of freedom, each of which encodes the decision of an agent to buy or sell a stock, and thus provides a natural expression to extract information for the opinion of the majority or minority. We therefore consider a novel term $h$, analogous to the one considered in the case of the Ising model~\citep{bornholdt}, but suitable for the continuous degrees of freedom of the $\phi^{4}$ theory:
\begin{equation}\label{eq:bor}
h(\phi_{i})=-a\phi_{i}\Bigg|\frac{1}{V} \sum_{i} \phi_{i} \Bigg|,
\end{equation}

The term $h$ therefore couples a fictitious field $a>0$ to the absolute value of the intensive magnetization.  We incorporate the term $h(\phi_{i})$ in the conditional probability distribution $p(\phi_{i}|\phi_{j,j\in n_{i}})$. The equations used to sample computationally the system remain identical excluding the value of the mean $\mu$ in Eq.~\ref{eq:mu} which is revised as:
\begin{equation}
\mu=\frac{c_{1}+h}{2c_{2}}.
\end{equation}

We recall that the field $\phi_{i}$ is not a binary degree of freedom but instead a continuous value $-\infty< \phi_{i}<\infty$.  The sign of the field ${sgn}(\phi_{i})$ then conveys the decision of an agent to buy or sell a stock and the magnitude of the field $|\phi_{i}|$ can be interpreted as a weight by which this decision is finalized. If the decision of an agent to buy or sell a stock, namely ${sgn}(\phi_{i})$, is therefore contingent on a price expectation, the magnitude of the field $|\phi_{i}|$ should therefore dictate how strongly the agent is convinced that the price is going to increase or decrease.  Consequently, the $\phi^{4}$ theory is capable of encoding an explicit variance in the expectation of agents within financial markets in contrast to spin models which consider binary or discrete degrees of freedom.   

Finally, we remark that the inclusion of the term $h(\phi_{i})$ of Eq.~\ref{eq:bor} can  induce an anti-ferromagnetic interaction, giving rise to a $\phi^{4}$ theory with frustrated ferromagnetic and anti-ferromagnetic dynamics. When the absolute value of the magnetization becomes sufficiently large, the term $h(\phi_{i})$ begins to dominate, thus forcing $\phi_{i}$ to change its sign. Consequently, a different physical behavior is expected for the multi-agent $\phi^{4}$ theory in relation to the conventional system.   

\section{\label{sec:level4} Stylized facts of $\phi^{4}$ financial markets}
We now implement Gibbs sampling for values $m^{2}=-3.0$, $\lambda=0.7$ and $a=5.0$ to obtain, after equilibration, $10^{5}$ configurations of the multi-agent $\phi^{4}$ model of financial markets for lattice size $L=64$. We remark that the specific values of $m^{2}=-3.0$, $\lambda=0.7$ define a $\phi^{4}$ theory, when excluding the introduced term $h$, which resides in the broken-symmetry phase~\citep{PhysRevD.79.056008,PhysRevLett.128.081603,arxiv.2205.08156}.  Our aim is to confirm that the model can reproduce behavior observed on empirical data. We therefore utilize as empirical data closing values from the FTSE 100 London Stock Exchange index, starting from the date 1/10/1984, and considering a daily increment until 30/09/2024.

\begin{figure}[t]
\includegraphics[width=8.6cm]{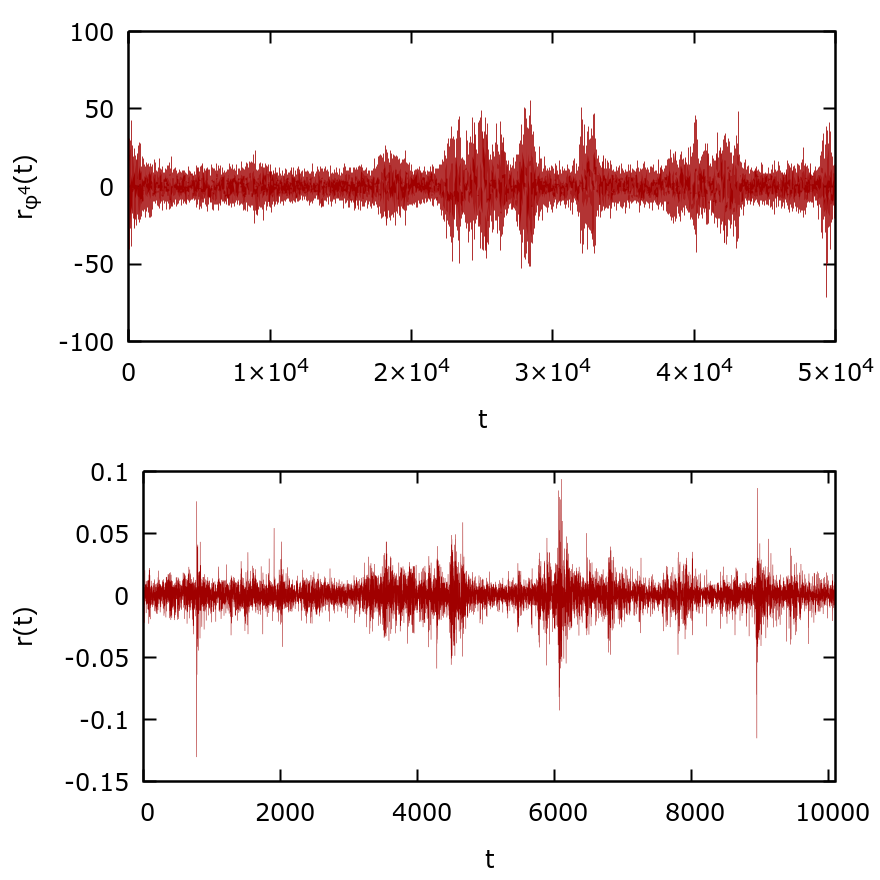}
\caption{\label{fig:fig2}  Returns $r_{\phi^{4}}(t)$ as obtained from the multi-agent $\phi^{4}$ theory (top) and log-returns $r(t)$ from the FTSE 100 index (bottom) versus time $t$. We observe on both cases the presence of intermittent phases with large fluctuations.  }
\end{figure}

Configurations of the multi-agent $\phi^{4}$ theory, interpreted as a model to represent expectations of agents in financial markets, are depicted in Fig.~\ref{fig:fig1}. Specifically, we depict configurations after $20500$, $28311$, and $30000$ Monte Carlo sweeps. A set of initial configurations has been discarded for equilibration. We observe a metastable state for $20500$ sweeps where global structure is evident, and a transition to a turbulent phase of $28311$ sweeps where the global structure deteriorates. The system is subsequently undergoing consecutive phase transitions as evident from the re-emergence of a metastable state for $30000$ sweeps. We remark that metastable states are interpretable as speculative bubbles of a stock that emerged without a fundamental cause and phases of chaotic dynamics indicate time frames where chartists dominate.

Empirical studies on financial data consider quantities such as log-returns $r(t)$ of stock prices $p(t)$ in relation to time $t$,
\begin{equation}
r(t)=\frac{\ln p(t)}{\ln p(t-1)}.
\end{equation}

In the case of models of financial markets derived by physical systems such as spin models one can interpret the difference in the magnetization as a measure of returns~\citep{KAIZOJI2002441}:
\begin{equation}
r_{\phi^{4}}(t)=\frac{M(t)-M(t-1)}{2}.
\end{equation}

\begin{figure}[t]
\includegraphics[width=8.6cm]{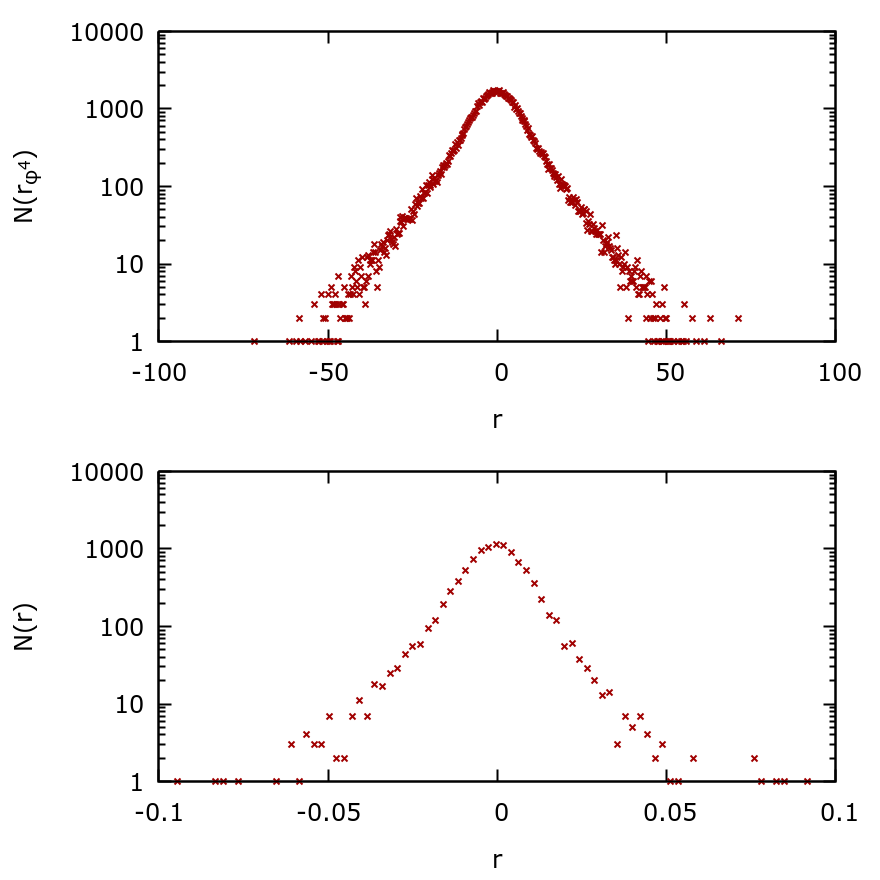}
\caption{\label{fig:fig3} Histograms of returns $r_{\phi^{4}}(t)$ as obtained from the multi-agent $\phi^{4}$ theory (top) and histograms of log-returns $r(t)$ as obtained from the FTSE 100 index (bottom) versus the value of returns. We observe the emergence of fat-tailed probability distributions. The y axis is logarithmic.}
\end{figure}

The returns $r_{\phi^{4}}(t)$ of stock prices as calculated on configurations of the $\phi^{4}$ theory are depicted on Fig.~\ref{fig:fig2}. In addition we include for a comparison the logarithmic returns $r(t)$ on the FTSE 100 index. We observe for both $r_{\phi^{4}}(t)$ and $r(t)$ intermittent phases with large fluctuations. The presence of intermittent phases on the multi-agent $\phi^{4}$ theory is a first indication that the system has a sufficient representational capacity to be utilized also for quantitative studies of empirical financial data.

A stylized fact of empirical financial data is the emergence of fat-tailed distributions of returns due to the high probability of extreme values to occur on real data. The histograms for both $r_{\phi^{4}}(t)$ and $r(t)$ are depicted in Fig.~\ref{fig:fig3}. We observe fat-tails on the histograms of returns for the multi-agent $\phi^{4}$ theory indicating a deviation from the Gaussian distribution. In addition we depict the empirical FTSE 100 London Stock Exchange data in which the aforementioned observation can be compared. The presence of fat-tailed distributions can be quantitatively assessed via the calculation of the excess kurtosis~\citep{PAGAN199615}:
\begin{equation}
\kappa= \frac{\mathbb{E}[(X-\mathbb{E}[X])^{4}]}{\sigma^{4}(X)}-3,
\end{equation}
where $\mathbb{E}[(X-\mathbb{E}[X])^{4}]$ is the fourth central moment of a random variable $X$ and $\sigma(X)$ is the standard deviation. We therefore calculate the excess kurtosis using the values of returns from the entire dataset and obtain $\kappa_{\phi^{4}}=3.37$ and $\kappa_{FTSE}=10.42$, indicating a deviation from the Gaussian distribution for which $\kappa_{Gaussian}=0$. We remark that $\kappa_{\phi^{4}}$ can fluctuate based on the sample size due to the presence of intermittent phases and should be calculated in principle for an asymptotically large number of Monte Carlo samples. In addition, the excess kurtosis is dependent on the choice of couplings $m^{2}$, $\lambda$, and $a$. We remark that empirical data on daily returns are anticipated to yield values of $\kappa$ in the range $2$ to $50$~\citep{PAGAN199615}.

\begin{figure}[t]
\includegraphics[width=8.6cm]{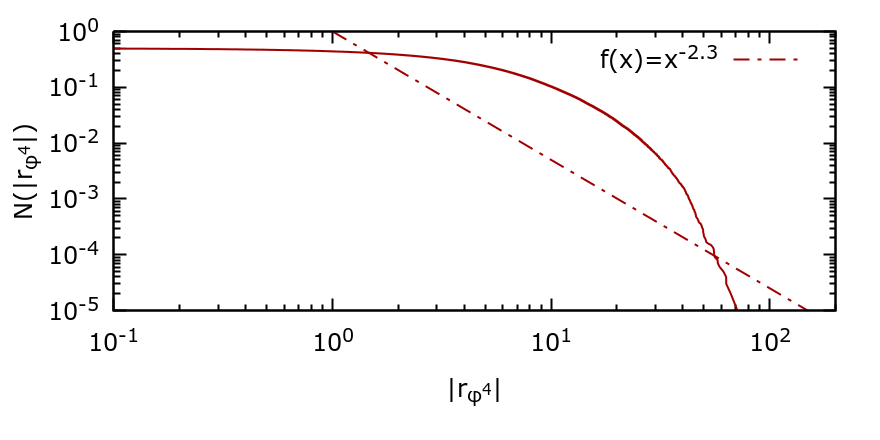}
\caption{\label{fig:fig4} Complementary cumulative distribution histograms of the absolute value $|r_{\phi^{4}}(t)|$ on logarithmic scales. Both axes are logarithmic.  }
\end{figure}

Useful information can be additionally extracted from the cumulative distribution function of absolute returns $|r_{\phi^{4}}(t)|$. In Fig.~\ref{fig:fig4} we depict the complementary cumulative distribution function of $|r_{\phi^{4}}(t)|$. We remark that empirical studies on financial data verify the presence of power-law scaling behavior $P(|r(t)|>x)\sim 1/x^{\theta} $ in the values of absolute returns $|r(t)|$ based on a critical exponent in the range $2\leq \theta \leq 4$. Spin models of financial markets claim consistent results for a value of $\theta=2.3$~\citep{KAIZOJI2002441}, and we include for comparison in Fig.~\ref{fig:fig4} a line proportional to the case of the exponent $\theta=2.3$.

\begin{figure}[t]
\includegraphics[width=8.6cm]{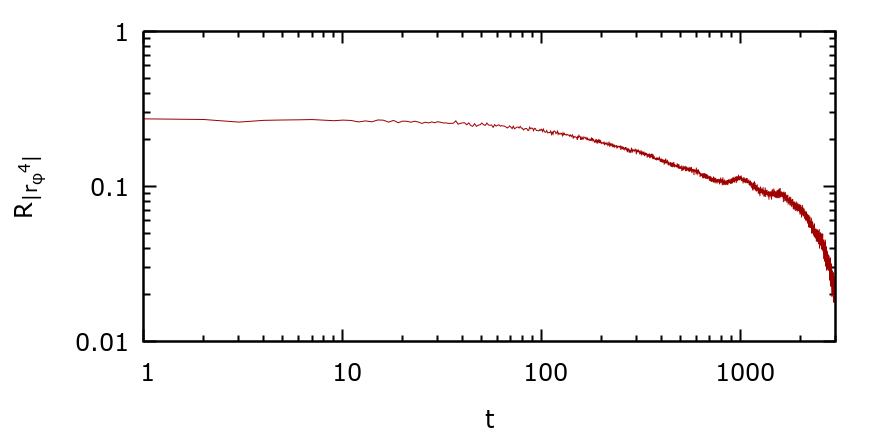}
\caption{\label{fig:fig5} Autocorrelation function for the absolute returns $|r_{\phi^{4}}|(t)$ as calculated on the multi-agent $\phi^{4}$ theory. Both axes are logarithmic. }
\end{figure}

Another observation on empirical data is the presence of clustered volatility where absolute values of returns $|r(t)|$ remain positively correlated for extended  periods of time and power-law decay of autocorrelations can be observed. This observation can be already inferred as valid from Fig.~\ref{fig:fig2}, where clustered phases have emerged. In Fig.~\ref{fig:fig5} we depict the autocorrelation function for $|r_{\phi^{4}}(t)|$  obtained from the multi-agent $\phi^{4}$ theory, where similar autocorrelation scales to empirical data are observed. The above observations therefore verify that the $\phi^{4}$ lattice field theory can reproduce nontrivial aspects of financial markets and can potentially be extended to model empirical financial data.

\section{\label{sec:level5}Conclusions} In this manuscript we introduced a $\phi^{4}$ lattice field theory with frustrated dynamics as a multi-agent system to reproduce stylized facts of financial markets such as fat-tailed distributions of returns and clustered volatility. Specifically, each lattice site, represented by a continuous degree of freedom, corresponds to an agent aiming to buy or sell a stock in a model of financial markets where expectation bubbles and crashes emerge. Two types of agents are represented in this model. Fundamentalists, which are guided by the fundamental value of an asset when forming decisions, and chartists which are guided instead by emerging trends. 

Using the multi-agent $\phi^{4}$ theory we demonstrated that we can reproduce, on a qualitative level, behavior observed on empirical financial data from the FTSE 100 London Stock Exchange index. This implies that the system possesses sufficient representational capacity to be extended on practical quantitative research problems. For example, one could solve the inverse problem~\citep{PhysRevD.103.074510} by searching for the most accurate values of the couplings $m^{2}$ and $\lambda$, defining a system in the broken symmetry phase, and a value of the field $a$, which are able to reproduce observables from empirical financial data. In addition, to further increase the representational capacity of the model and allow for an additional set of variational parameters to be learned, one could consider the couplings $m^{2}$ and $\lambda$ to be inhomogeneous. Of additional interest would be to consider a subset of learnable interactions $J_{ij}$ between agents to be disordered, giving rise to a system which would partially resemble a $\phi^{4}$ spin glass~\citep{phi4glass}. 

The $\phi^{4}$ lattice field theory is a system with continuous degrees of freedom, namely $-\infty < \phi <\infty$. We discussed how we can interpret the presence of continuous degrees of freedom as a magnitude that weighs the decision of an agent to buy or sell a stock. The multi-agent $\phi^{4}$ theory therefore provides a different perspective to the representation of agent expectations in relation to systems with binary or discrete degrees of freedom. We remark that the $\phi^{4}$ theory is a generalization of the Ising model since the latter can be recovered as a limiting case. The $\phi^{4}$ phase transition is additionally in the same universality class as that of the Ising model, and the system includes a crossover to the Gaussian case for $\lambda=0$. The aforementioned observations imply that the $\phi^{4}$ theory is expected to be equally successful as the Ising model in pertinent problems, with the additional benefit of being able to represent cases which necessitate the consideration of continuous degrees of freedom. Given the success of the Ising model in artificial intelligence research of multi-agent systems, we expect that the $\phi^{4}$ theory could be extended to applications that reach far beyond the representation of simplified models of financial markets.  

\section{\label{sec:level6}Data Availability Statement} Code to simulate the multi-agent $\phi^{4}$ theory is available on Ref.~\footnote{\href{https://github.com/dbachtis/phifm}{https://github.com/dbachtis/phifm}}. Data to reproduce the figures are available from the author upon request.

\section{\label{sec:level7}Acknowledgements} This project has not received any commercial funding.  The author acknowledges support from the Science and Technology Facilities
Council (STFC) Consolidated Grant ST/X00063X/1 ``Amplitudes, Strings \& Duality".

\bibliography{ms}

\end{document}